\documentclass[aps,prl,twocolumn]{revtex4}
\usepackage{epsfig}

\begin{document}

\title{
Coherent population trapping in ruby crystal at room temperature
}

\author{
	Roman Kolesov
}

\address{
	    	Department of Physics and Institute for Quantum Studies,
                          Texas A\&M University,\\
		College Station, Texas~~77843-4242\\
}
\date{\today}

\begin{abstract}
Observation of coherent population trapping (CPT) at ground-state Zeeman sublevels
of $Cr^{3+}$-ion in ruby is reported. The experiments are performed
at room temperature by using both nanosecond optical pulses and nanosecond trains
of ultrashort pulses. In both cases sharp drops in the resonantly induced
fluorescence are detected as the external magnetic field is varied.
Theoretical analysis of CPT in a transient regime due to pulsed action of optical
pulses is presented.
\end{abstract}
\maketitle

In last few decades very much attention has been paid to studying optical interference phenomena
originating from the atomic coherence excited
by laser light while interacting with multi-level atomic media. These
phenomena include coherent population trapping (CPT) \cite{cpt_ref}, electromagnetically
induced transparency (EIT) \cite{eit_ref}, lasing without population inversion (LWI)
\cite{lwi_ref}, slow light \cite{slow}, and many others. The interest to this field
arises from a number of potential applications of these phenomena. Among
them are metrology, quantum information storage and processing, ultra-sensitive
magnetometry, development of lasers in vacuum ultraviolet (VUV), X-ray, and $\gamma$-ray
ranges, etc. However, CPT and related phenomena are mostly being studied in gaseous media
while for practical purposes solid-state materials are preferable. The advantages of solids
are high atomic density, absence of atomic diffusion, compactness, and robustness. At the
same time, decoherence processes in solid media are severe. Nevertheless, several
demonstrations of EIT \cite{Pr_YSO,EIT}, slow light \cite{Pr_YSO},
and non-degenerate four-wave mixing \cite{hakuta} in solids have been reported by several groups.
However, all those works were performed at temperatures close to the liquid helium one. This fact,
in turn, makes real solid-state applications of the above phenomena questionable.

In this Letter I report the first, to my knowledge, experimental observation
of CPT resonances in crystalline solid at room temperature. Prior to that, I have to
discuss some general issues in order to identify the class of solid-state materials
in which the phenomena of EIT/CPT can be observed at room temperature. I will consider only
optical crystals doped with either rare-earth or transition metal ions possessing
discrete electronic states in the bandgap of the host
material. The electronic transitions between those states, both optical and Zeeman and/or
hyperfine, can be very narrow even at room temperature. The class of so-called S-state
ions doped into wide-bandgap optical crystals is of particular interest. It includes
$Cr^{3+}$, $Eu^{2+}$, $Fe^{3+}$, $Mn^{4+}$, $Gd^{3+}$, and several others. The most important
feature of these ions is that, when incorporated into a crystal, their first excited electronic
state lies several thousands $cm^{-1}$ above the ground one. Consequently,
the strongest relaxation process between the ground-state Zeeman or hyperfine sublevels,
owing to resonant inelastic scattering of phonons at low-energy electronic transition
(Orbach relaxation, \cite{Orbach}) is suppressed. As a result, the decoherence rate
at the transitions between Zeeman/hyperfine sublevels can be as low as a few $MHz$
even at room temperature. This fact is confirmed by a vast number of electron paramagnetic
resonance (EPR) measurements in which the ions mentioned above serve as paramagnetic probes.
Some of them, namely, $Cr^{3+}$, $Eu^{2+}$, and $Gd^{3+}$, offer very favorable optical
properties as well. In the following discussion, the case of ruby, i.e. $Cr^{3+}:Al_2O_3$,
is considered.

Another important feature of doped solids is that the population decay from
the excited electronic state is much slower than decoherence at Zeeman/hyperfine
transition. This is exactly opposite to the case of alkali-metal
vapors in which most of experimental work on EIT/CPT is being performed. Therefore, one cannot
expect to populate the coherent superposition of ground-state sublevels, which
does not interact with the laser field under the conditions of CPT (so-called, ``dark'' state),
by decay of the excited optical level, as is the case in the alkali-metal experiments.
The only way of exciting ground-state coherence in that situation is to remove as much population
as possible from the ``bright'' state (the coherent superposition of sublevels orthogonal to the
``dark'' one) by transferring it to the upper optical level. This should be done faster
than the coherence decays, i.e. laser excitation has to be pulsed.

\begin{figure}
\center{
\includegraphics[width=6cm]{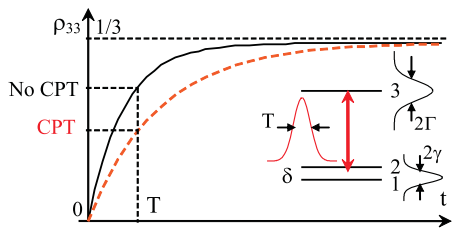}
\caption{\label{fig:growth} Temporal evolution of the upper optical state population
under pulsed excitation. The inset shows the model of a three-level atomic medium interacting
with light pulse.}
}
\end{figure}

Let us consider the following theoretical model. A three-level medium
is illuminated with a resonant laser pulse interacting with
both $1\leftrightarrow 3$ and $2\leftrightarrow 3$ optical transitions (see inset of Fig.\ref{fig:growth}).
In the rest of the paper we will be interested in the amount of population transferred into
level $3$ by a laser pulse or pulse train as a function of the energy separation between ground-state Zeeman sublevels
$1$ and $2$ which can be varied by means of external magnetic field. The following notations are introduced: $\Gamma$
is the homogeneous halfwidth of optical transitions, $\gamma$ is the Zeeman decoherence rate,
$\Omega$ is the laser field time-dependent Rabi frequency (dipole matrix elements of both optical
transitions are assumed equal). The initial populations of the two ground-state sublevels
are $\rho_{11}=\rho_{22}=1/2$, $\rho_{33}=0$. Assume also, that the pulse duration is longer
than $\Gamma^{-1},\gamma^{-1}$. This assumption somewhat contradicts the last statement
in the previous paragraph, but it is relevant under the experimental conditions described below.
Algebraically solving the density matrix equations for all the coherences, one comes to
the following linear equation for the upper state population:
\begin{equation}
\label{eq:pop3}
\frac{\rho_{33}}{dt}=P\left(1-\frac{P\left(P+\gamma\right)}{\delta^2+\left(P+\gamma\right)^2}\right)
\left(1-3\rho_{33}\right),
\end{equation}
where $P=2\Omega^2/\Gamma$ and $\delta$ is the frequency separation between the ground-state
sublevels. The second term in brackets represents the effect of Zeeman coherence on the
rate at which level $3$ is populated. Laser pulse starts acting on the medium at $t=0$.
As $t\rightarrow\infty$, all three levels become equally populated, as expected. However,
if at some instant $T$ the laser pulse is shut off, the final population of level $3$ is
smaller for $\delta=0$ than it is for $\delta\neq 0$ because of its lower growth rate
as illustrated in Fig.\ref{fig:growth}. The width of this transient CPT resonance
is determined by $P+\gamma$, i.e. by Zeeman decoherence rate and power broadening.

Similar effects can be observed under the illumination of the medium with a short (nanosecond)
train of ultrashort (picosecond) pulses. In addition to CPT resonance at zero magnetic field,
one should expect drops in the upper state population as the Zeeman sublevel separation
becomes equal to the multiple of the pulse repetition rate in the train \cite{train_kochar}.
It worths mentioning that transient coherent effects in ruby crystal at room temperature
were first studied in 1974 \cite{magnetization}. In that work precessing magnetization
was induced by a train of ultrashort pulses. Significant enhancement in magnetization was
observed as the splitting between Zeeman sublevels of $Cr^{3+}$-ion became multiple of
the pulse repetiotion rate. However, no effect of Zeeman coherence on the absorptive
properties of the crystal was studied.

\begin{figure}
\center{
\includegraphics[width=5cm]{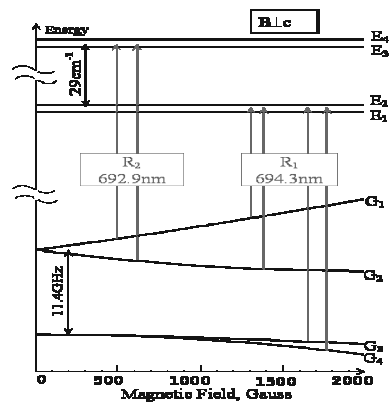}
\caption{\label{fig:levels} Energy levels of $Cr^{3+}$-ion. Optical transitions allowed for
$\sigma$-polarization in ${\bf B}\perp c$ geometry are indicated.}
}
\end{figure}

To understand how the above theoretical description applies to ruby, let us consider
the energy levels of $Cr^{3+}$ in that crystal (see Fig.\ref{fig:levels}).
Its ground state is described by the following spin Hamiltonian:
\begin{equation}
\label{eq:hamiltonian}
H=g_{||}B_{||}S_z+g_{\perp}{\bf B}_{\perp}{\bf S}_{\perp}+D\left(S_z^2-S\left(S+1\right)/3\right),
\end{equation}
where $g_{||}=1.982$ and $g_{\perp}=1.987$ are the ground-state $g$-factors in the directions parallel
and perpendicular to the crystal $c$-axis respectively, $B_{||}$ and ${\bf B}_{\perp}$ are
the corresponding magnetic field components, $2D=-11.47\;GHz$ is the zero-field
splitting of the ground state, and $S=3/2$ is the spin of $Cr^{3+}$. The excited
optical state is split into two spin-doublets separated by $29\;cm^{-1}$ giving rise
to $R_1$ ($694.3\;nm$) and $R_2$ ($692.9\;nm$) optical lines of ruby. The width of both
optical transitions is $11\;cm^{-1}$ while the oscillator strength is $7.5\times 10^{-7}$.

The situation when ${\bf B}\perp c$-axis is of particular interest. The selection rules for
$\sigma^{\pm}$-polarizations of light propagating along the c-axis are summarized in Tables
\ref{tab:selection1} and \ref{tab:selection2} \cite{selection}. The matrix elements are normalized so that
the sum of their squares for each of the $\sigma^{\pm}$-polarizations is unity. One can easily see that there are a number
of Raman processes connecting levels $G_1$ and $G_2$ and $G_3$ and $G_4$. Since
the optical linewidth is greater than the separation between the ground-state Zeeman sublevels,
all $\Lambda$-systems will give their contributions to the Zeeman coherence excitation.
Therefore, if one uses a linearly polarized laser light, the net contribution of all
Raman processes to the coherence excitation at either $G_1\leftrightarrow G_2$ or
$G_3\leftrightarrow G_4$ transition is exactly zero. For example, at $R_2$-line the coherence
between levels $G_1$ and $G_2$ due to Raman process going though $E_3$ for $\sigma^{+}$-polarization
is exactly compensated by the one going though $E_4$ for $\sigma^{-}$-polarization.
The situation is the same for $R_1$-line. However, there is no such difficulty for circularly polarized light.
\begin{table}[t]
\begin{center}
\begin{tabular}{|c|cccc|}
\hline
State	& $G_1$		& $G_2$		& $G_3$		& $G_4$		\\
\hline
$E_1$	& $\pm 0.25$	& $0.25$	& $-0.31$ 	& $\pm 0.31$	\\
$E_2$	& $0.25$	& $\pm 0.25$	& $\pm 0.31$	& $-0.31$	\\

\hline
\end{tabular}
\end{center}
\caption{\label{tab:selection1}Selection rules for $\sigma^{+}$ (upper sign) and
$\sigma^{-}$ (lower sign) polarizations interacting with $R_1$-line.}
\end{table}\begin{table}[t]
\begin{center}
\begin{tabular}{|c|cc|cc|}
\hline
State	& \multicolumn{2}{|c|}{$G_1$}		& \multicolumn{2}{|c|}{$G_2$}		\\
	& $\sigma^{+}$ & $\sigma^{-}$ 		& $\sigma^{+}$ & $\sigma^{-}$		\\
\hline
$E_3$	& $0.43$	& $-0.07$		& $-0.43$	& $-0.07$		\\
$E_4$	& $-0.07$	& $-0.43$		& $0.07$	& $-0.43$		\\

\hline
\end{tabular}
\end{center}
\caption{\label{tab:selection2}Selection rules for $\sigma^{\pm}$ polarizations interacting with $R_2$-line. Transitions from
the lower ground-state doublet to levels $E_3$ and $E_4$ are forbidden.}
\end{table}

\begin{figure}
\center{
\includegraphics[width=8cm]{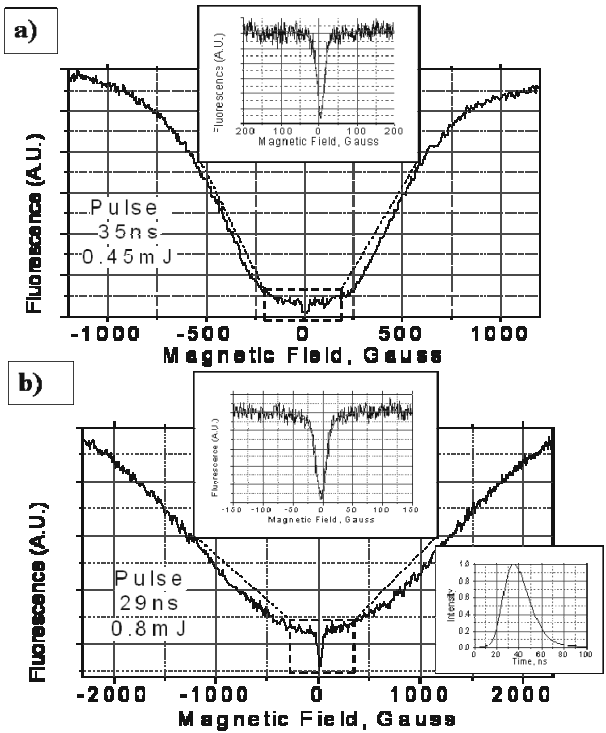}
\caption{\label{fig:long_exp} CPT resonances in zero magnetic field: a) trace for $R_1$-line and b) trace for $R_2$-line.
Narrow peaks are shown in detail on insets. See text for details.}
}
\end{figure}

\begin{figure}
\center{
\includegraphics[width=8cm]{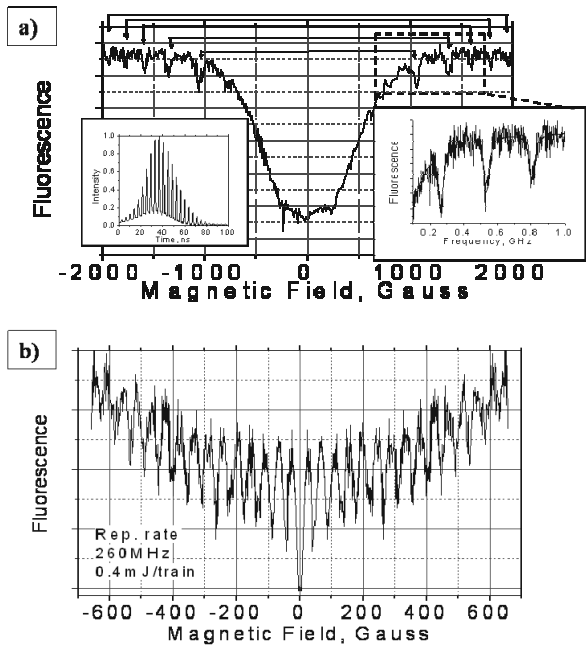}
\caption{\label{fig:train_exp} CPT resonances with mode-locked laser: a) trace for $R_1$-line and b) trace for $R_2$-line.
See text for details.}
}
\end{figure}

The experiments were performed on a $3\times 3\times 5\; mm^3$ dilute ruby crystal
($0.002\%$ of $Cr^{3+}$) with $c$-axis being perpendicular to one of $3\times 5\;mm^2$ sides
(Scientific Materials Corp.) A homemade tunable Ti:Sapphire laser pumped
by the doubled $Nd:YLF$ pulsed laser (Photonics Industries, model GM-30) could operate
both in a long pulse regime (pulse duration $\approx 30\;ns$ FWHM) and in a mode-locked regime
delivering $\approx 30\;ns$ pulse trains with repetition rate $\approx 260\;MHz$.
The laser was mode-locked by using a cryptocyanine saturable absorber dissolved
in ethanol. The typical
output energy per pulse or per train was several tenths of a $mJ$ at a $100\;Hz$ repetition rate. The laser polarization
was linear. The typical shapes of pulses in both regimes are shown on the insets of
Figs.\ref{fig:long_exp} and \ref{fig:train_exp}. The shape of the pulse train
was taken with a $350\;ps$-risetime photodiode and a $500\;MHz$-bandwidth
oscilloscope. The duration of
each individual pulse in the train was not measured.
The crystal was placed into a varying magnetic field with ${\bf B}\perp c$-axis and
illuminated by laser pulses sent through
a $\lambda/4$-plate and focused by an $f=10\;cm$ lens. Presumably, the laser spot diameter
at the sample was $\sim 100\;\mu m$. The fluorescence was detected as a function
of slowly varying periodic magnetic field. The sweeping frequency of the magnetic field was $0.17\;Hz$.


The experimental results for the case of long pulse excitation for both $R_1$ and $R_2$ lines
are shown in Fig.\ref{fig:long_exp}. The spectra were averaged over several hundreds of magnetic field sweeps.
In the case of $R_1$-line, there are two features present in the spectrum:
the broad one with the plateau on the bottom and a sharp one of much smaller amplitude. The smaller one
corresponds to less than $0.5\%$ decrease in the fluorescence amplitude and has a width of $18\;G$. It corresponds
to CPT resonance due to $G_1\leftrightarrow E_1,E_2 \leftrightarrow G_2$ $\Lambda$-system, thus its width
in terms of frequency units is $\approx 100\;MHz$ since $g$-factor for $G_1\leftrightarrow G_2$ transition
is $\approx 4$. The broad drop in fluorescence signal has an amplitude $\approx 7-8\%$. Its FWHM is close to
$\approx 950\;G$. The presence of plateau at the bottom of that feature indicates that this is a CPT resonance
in the $G_3\leftrightarrow E_1,E_2 \leftrightarrow G_4$ $\Lambda$-system with levels $G_3$ and $G_4$
being split very slowly as the magnetic field increases. Its width recalculated into frequency
domain is $\approx 40-50\;MHz$, i.e. this CPT resonance is, in fact, much sharper than the one discussed
above. The possible explanation for this fact is as follows. The main contribution to the EPR linewidth
of $Cr^{3+}$ in ruby is known to originate from magnetic dipole-dipole hyperfine interaction
with neighboring $^{27}Al$ nuclei \cite{broadening}. Thus, the magnetic-allowed Zeeman transition
$G_1\leftrightarrow G_2$ should be broadened much stronger than the magnetic-forbidden $G_3\leftrightarrow G_4$.
This also explains more than an order of magnitude difference in the amplitudes of those two CPT resonances, since
in $\Lambda$-system with longer-lived Zeeman coherence CPT should be more pronounced.

For the $R_2$-line only one CPT resonance, represented by a sharp feature at the bottom of a smooth
background, is observed. It has the width of $18\;G$, i.e. the same as the one observed at $R_1$-line,
thus it corresponds to $G_1\leftrightarrow E_3,E_4 \leftrightarrow G_2$ $\Lambda$-system. The decrease
in the fluorescence is $3-4\%$ at the center of the resonance. The fact that this resonance is much stronger
is easily understandable, since 1) there is no background fluorescence excited from the other pair of
levels, $G_3$ and $G_4$, and 2) the transitions $G_1,G_2\leftrightarrow E_3,E_4$ are 3 times stronger
than $G_1,G_2\leftrightarrow E_1,E_2$. The smooth background originates, probably, from level
mixing in high magnetic fields and consequent change in the optical selection rules. That was confirmed
by removing the $\lambda/4$-plate, i.e. by making laser light linearly polarized. Under this condition all
three CPT resonances, discussed above, disappeared, as they were supposed to, while the smooth feature in question remained.

The results for trains of ultrashort pulses are shown in Fig.\ref{fig:train_exp}. In the spectrum obtained for
$R_1$-line additional resonances appear in pairs. They are not equidistant because of nonlinear dependence of the
levels $G_3$ and $G_4$ separation on the magnetic field strength. However, if recalculated into
frequency domain, their positions exactly correspond to the harmonics of laser repetition rate ($260\;MHz$).
The first three of them are shown on the inset of Fig.\ref{fig:train_exp}a. Their widths (FWHM) are in the range $35-38\;MHz$
and their positions, according to the fit, are $265\;MHz$, $533\;MHz$, and $804\;MHz$. The widths are in
rather good agreement with the result obtained for the $G_3\leftrightarrow G_4$ transition
in a long pulse regime. It was not possible to observe multiple CPT resonances corresponding to
$G_1\leftrightarrow E_3,E_4 \leftrightarrow G_2$ $\Lambda$-system with appreciable signal-to-noise ratio
due to very small amplitude of the CPT effect ($<0.5\%$) and rather unstable operation of laser
in a mode-locked regime. However, the signature of these resonances is the increased quasi-noise
at the bottom of the broad feature.

For $R_2$-line almost 30 CPT resonances are indicated. In order to improve the signal-to-noise ratio, the original
spectrum was flipped with respect to the zero magnetic field and the two sets of data were combined together.
That explains the fact that the plot shown in Fig.\ref{fig:train_exp}b is completely symmetric. The magnetic field
separation between each pair of resonances is $\approx 45\;G$. This value corresponds to $252\;MHz$ of frequency
separation which is somewhat close to $260\;MHz$ of laser repetition rate. Due to rather poor signal-to-noise
ratio and rater low spectral resolution it is not possible to determine the linewidth of resonances exactly,
but it seems to correspond rather well to the value of $18\;G$ obtained in a long pulse regime.

In summary, the first observation of CPT in a crystal at room temperature is reported. Possible applications of
the reported phenomena include room-temperature all-optical analogue of EPR spectroscopy in low magnetic fields,
suppression of excited state absorption in laser crystals \cite{ESA}, alternative technique of mode-locking
in lasers \cite{mode_lock}, etc.

The author is grateful to F.Vagizov, P.Hemmer, Y.Rostovtsev, E.Kuznetsova, and O.Kocharovskaya, for stimulating discussions and to
S.Lissotchenko for technical assistance. The work was supported by NSF and AFOSR.

\end{document}